\documentclass{article}
\usepackage[utf8]{inputenc}
\usepackage{multirow}
\usepackage{graphicx}
\usepackage{booktabs}
\usepackage{multirow}
\title{Can We Detect Mastitis earlier than Farmers?}
\author{Cathal Ryan, Christophe Gu\'{e}ret, Donagh Berry, Brian Mac Namee}

\usepackage{graphicx}
\usepackage[official]{eurosym}
\usepackage{textcomp}
\usepackage{babel}
\usepackage{fourier} 
\usepackage{array}
\usepackage{makecell}

\begin{document}

\maketitle

\begin{abstract}
    The aim of this study was to build a modelling framework that would allow us to be able to detect mastitis infections before they would normally be found by farmers through the introduction of machine learning techniques. In the making of this we created two different modelling framework's, one that works on the premise of detecting Sub Clinical mastitis infections at one Somatic Cell Count recording in advance called SMA and the other tries to detect both Sub Clinical mastitis infections aswell as Clinical mastitis infections at any time the cow is milked called AMA. We also introduce the idea of two different feature sets for our study, these represent different characteristics that should be taken into account when detecting infections, these were the idea of a cow differing to a farm mean and also trends in the lactation. We reported that the results for SMA are better than those created by AMA for Sub Clinical infections yet it has the significant disadvantage of only being able to classify Sub Clinical infections due to how we recorded Sub Clinical infections as being any time a Somatic Cell Count measurement went above a certain threshold where as CM could appear at any stage of lactation. Thus in some cases the lower accuracy values for AMA might in fact be more beneficial to farmers.
\end{abstract}
\section{Introduction}
Mastitis is one of the most common infections that harm dairy farms around the world today with approximately 20-30\% of a herd being affected around the world \cite{heringstad_klemetsdal_ruane_2000} with it resulting in many different effects such as reduced milk yield, veterinary costs, and increased risk of early culling \cite{cavero}. Mastitis infections can be further split into two seperate categories mainly Sub Clinical Mastitis Infections (SCM) and also Clinical Infections (CM). They are mainly split on how they can be found by farmers or veterinarians, CM infections are mastitis infections which create swelling or \linebreak{} clotting in the udder while SCM are seen as any mastitis infection that is only found through the measuring of Somatic Cell Count (SCC) or the use of California Mastitis Test.\par 
Many new decision support systems have been reported to help try to reduce the effect of mastitis incidence or to detect these infections using different approaches. This can be broken into the idea of detecting infected cows compared to healthy cows \cite{hyde_down_bradley_breen_hudson_leach_green_2020} or the detection of individual infections \cite{friggens_chagunda_bjerring_ridder_hojsgaard_larsen_2007}, \cite{ebrahimie_ebrahimi_ebrahimi_tomlinson_petrovski_2018}, \cite{mol_ouweltjes_2001}, etc. Many of these studies have certain downfalls such as them only including a small sample size of cows \cite{kim_min_choi_2019}, \cite{mol_ouweltjes_2001}, using the same herd \cite{ebrahimie_ebrahimi_ebrahimi_tomlinson_petrovski_2018}, \cite{chagunda_friggens_rasmussen_larsen_2006}, \cite{panchal_sawhney_sharma_dang_2016}, \cite{sun_samarasinghe_jago_2009} or the same year \cite{kamphuis_mollenhorst_feelders_pietersma_hogeveen_2010}, \cite{santman-berends_riekerink_sampimon_schaik_lam_2012} for the entire dataset. These three examples are ways in which a study might be able to achieve over exagerated results due to the model learning a pattern that might not be present in more general data. This was reduced in our study through the use of both multiple farms and years in our dataset while also including a large quantity of both clinical and sub clinical infections seen in Table \ref{tab:mast inf}.
\par Within this study we use a wide range of features that are widely available to farmers to help produce a modelling framework that could help reduce the impact of mastitis each year. We also try to improve on these base features through the introduction of two sets of different feature ideas, the first being the calculation of how different a cows feature is to the mean value for that farm while the second feature is the addition of using time series data preparation approaches to describe the changes of features over time during a lactation. With these two sets of features we set up two modelling frameworks that could be used to determine \linebreak{}mastitis infections before they occurred. The first modeling framework \linebreak{}(named Somatic Cell Count Milking Alert) contained days at which Somatic Cell Count (SCC) was recorded only and then we created a single alert that would \linebreak{} determine if there was a high enough probability that the cow would become infected with a Sub-Clinical Mastitis (SCM) infection the next time SCC was recorded, while the second modelling framework (named All Milking Alert) contained all days where a cow was milked and a new alert was created for each of these which determined if there was a high probability that the cow would become \linebreak{}infected with either a SCM infection or a Clinical Mastitis (CM) infection within the next 7 days. \par The contributions of this paper include the two different types of modelling \linebreak{} frameworks namely AMA and SMA which show that using data widely available on Irish Dairy Farms. The rest of the paper is as followed, in Section 2 we introduce the different aspects included in this study, in Section 3 we introduce the different results associated with our study, in Section 4 we interpret the outcome of our results and then we conclude with Section 5 with a conclusion.

\begin{table}[ht]
\begin{center}
    
\begin{tabular}{@{}cc@{}}
\toprule
Clinical & Sub Clinical \\ \midrule
2791     & 18175       \\ \bottomrule
\end{tabular}
\caption{Total Amount of Mastitis Infections}
\label{tab:mast inf}
\end{center}

\end{table}

\section{Materials And Methods}
\subsection{Data}
\begin{figure}[ht]
\centering
\includegraphics[scale = 0.25]{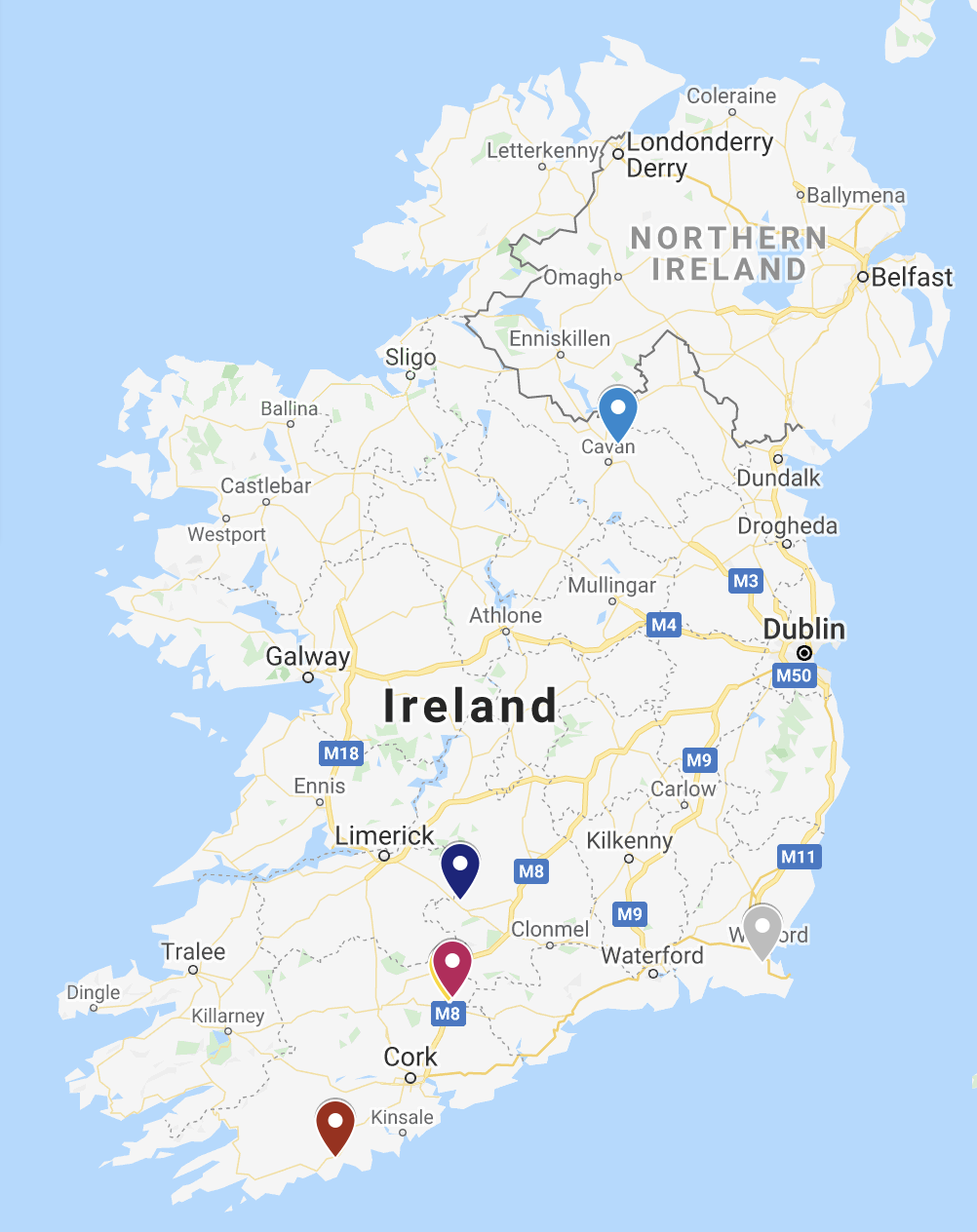}
\label{fig:MAP}
\end{figure}

The data used within this study comprised of milking and health data provided from Cows situated in 7 different research farms around Ireland which can be seen in Figure \ref{fig:MAP} where each of the pointers points to a group of farms, this is due to the fact that some farms are closer in distance to others. The data contained information on cows born after 1st January 2010 and contained many different aspects of the cows history such as the milking characteristics for the cow, treatment records for the cow, the dry off and calving dates for each lactation, the farm each cow was located on, information on the genetics of the cow where available, clinical infection records for the cow and also information on the weight of the individual cow. \par Due to the nature of how treatment administration could in some way affect the milking characteristic of a cow we decided to remove any milking observation that was recorded less than 7 days after an infection. To make sure we didn't use the same information when training and testing our model we decided to separate our combined data into two separate data sets, these being calving periods up to and including 2017 and calving periods within 2018.   

\subsection{Features}
The features used within this study comprised of both basic milking characteristics at a cow level, the amount of clinical and subclinical infections for that cow, general information about each milking the cow was recorded to have and the genetic merit for that cow expressed as PTA.
\subsubsection{Basic Milk Characteristics}
The data used for this study included milk recordings for each cow. Cows in this study were milked twice each day with different features being recorded at different intervals. Total yield and maximum flow rate were individually milked at each \linebreak{} milking every day that the cow was milked. The next most frequent recorded features were Fat \%, Protein \% and Lactose \% which were recorded every 7 days with their AM and PM recordings being measured at different days which is shown in Table \ref{fig:AMPM recordings}. The least frequently recorded feature was Somatic Cell count (SCC) which was only measured during AM milkings at times were Fat percentage, Protein percentage and Lactose were recorded. Due to the fact that not all the features were measured each time the cow was milked we chose to keep the last observed value within a particular lactation as the non recorded feature.

\begin{table}[ht]
\caption{AM \& PM recordings}
\label{fig:AMPM recordings}
\begin{tabular}{|l||lllllllllll|}
 \hline
Days in Milk  & 10  & 11 & 12 & 13  & 14 & 15 & 16 & 17  & 18 & 19 & 20\\  \hline
AM recordings & Yes & No & No & No  & No & No & No & Yes & No & No & No\\  \hline
AM fat & 4 & 4 & 4 & 4  & 4 & 4 & 4 & 4.5 & 4.5 & 4.5 & 4.5\\  \hline
PM recordings & No  & No & No & Yes & No & No & No & No  & No & No & Yes  \\  \hline
PM fat & 3.5 & 3.5 & 3.5 & 4  & 4 & 4 & 4 & 4 & 4 & 4 & 4.5\\  \hline
\end{tabular}
\end{table}

\subsubsection{Difference to Farm Averages}
One of the contributing aspects of this paper is the idea of if we can use how different a cow is to the farm average to determine if the cow is set to change from healthy to sick in the coming days. This idea was constructed on each of the basic features which led to 11 new features in our input features.
\subsubsection{Time Series Features}
These features used the concept of how a cow's milking characteristic changed\linebreak{} during its lactation period as a way of determining if the cow was set to change from healthy to sick in the coming days. For each milking feature we created its maximum, minimum, standard deviation, average and median value over the last 15 and 30 days. 

\subsubsection{Infection Features}
The data used for this study also included the dates at which a cow was said to have become infected with a CM or SCM infection. This information was used to create dummy features that could be seen as an illustration of how healthy the cow has been since it first started milking. This was done by counting up the total amount of Clinical or sub clinical infections for each cow for its entire milking history and also for each lactation separately. 
\subsubsection{General Information}
The data used for this study also included more basic information such as 'Days in Milk', 'Parity', 'Month'. Days in Milk corresponded to the amount of days from calving to that milking record, while Parity corresponded to that particular lactation value for the cow and finally Month corresponded to the Month each milking record was recorded in.
\subsubsection{Genetic Merit}
The data used also contained information regarding the genetic aspect of the cows history through the use of predictive transmitting ability (PTA) for each cow. Three different values for each cow were given. These were a value for the first lactation, second lactation and then one for any lactation after the second lactation.

\subsection{Model Building Approach}
There are some aspects of both models that are the same which includes removing milkings if they were less than 10 days since the cow had calved. This is talked about in many studies due to the idea that SCC is seen to be very unreliable for farmers until a certain period after calving, yet the amount of days at which is seen safe to use milkings is again not widely agreed upon. All other milkings were kept within the data sets for model 1 while for model 2 a further reduction was considered. This was in regards to any milking that occurred to soon after a previous infection of that type, the period chosen for this consideration was set to 7 days such that we could assume that the pattern formed by the infection was removed from the data and thus could be assumed to be again within a healthy state. We also created each model with using the features regarding infection history using both SCM and CM infections for both model types. The reason behind this is the idea that while CM infections are much less common than SCM and thus wont change frequently enough during a lactation or throughout a cows lifespan compared to SCM and thus SCM could possibly allow for larger subgroups in the data sets we are using.
\subsubsection{Somatic Cell Count Milking Alert}
Within this modelling framework we are trying to answer the question of if we can predict if a cow will be classed as having a SCM infection the next time they get recorded for SCC from the current SCC measurement and other relevant features. This as a result can only create alerts when SCC is being measured and thus would only give a farmer an alert for a small subset of days that their cows get milked and also is unable to check for CM infections using our current data preparation.
\subsubsection{Any Milking Alert}
This model uses all the data that is available instead of only milkings for which SCC was recorded. As a result the outcome of this model is quite significantly different to that of the first model. The first difference is that unlike the first model which can only classify SCM infections the second model can also classify CM due to the model creating a new alert at each time a cow was milked. \par The structure of our data for this modelling approach is illustrated in Table 3. From this table we can see this model type generates four new columns within our data. These being `Time Till Infection', `Time Healthy', `Early Detection' and also `Unsure Day'. These are calculated separately for clinical and sub clinical infections. `Time Till Infection' counts up the amount of days till the next infection of that type which if at a value between 1-7 is passed to the column `Early Detect' as a 1 and is seen as the positive class label when creating the model. While `Time Healthy' counts up the amount of days since the last infection which if at a value between 0-7 is passed to the column `Unsure Day' as a 1. Any milking day that has a value of 1 for `Unsure Day' is removed from the data set for that particular infection type. The results of this exclusion can be seen in Table \ref{tab: AMA structure}.

\begin{table}[ht]
\begin{center}
    
\begin{tabular}{@{}lllllll@{}}
\toprule
                 & Clinical  & Sub Clinical \\ \midrule
Milk Recordings & 11836    & 97430\\ \bottomrule
\end{tabular}
\caption{Amount of Milk Recordings Removed for Clinical and Sub Clinical Datasets}
\label{tab:Unsure Rows}
\end{center}

\end{table}

\par The reasons for selecting 7 days for the period of `Early Detect' was due to the fact that in our data SCC was measured on average once a week and thus similarities can be kept between this model and model 1 for SCM infections. While the reason for choosing the period for the `Unsure Days' was closely linked to the same idea but we decided to include milking days that were seen as the infected milking day such that even if we were only able to generate an alert on the last possible day it would still be atleast the day before the infection occured such that some precautions could be achieved within a farm. The addition of removing `Unsure Days' led us the possibility to have to remove a subset of a cows lactation period for longer than 7 days if the cows infection kept occuring which allowed us to make sure that any milkings \linebreak{} included in this model were either healthy milkings or milkings that could be \linebreak{} assumed to be not related to past infections.

\begin{table}[ht]
\begin{center}
\begin{tabular}{|l|l|l|l|l|l|}
\hline
\thead{Days \\ in Milk} & \thead{Infection \\ Recorded} & \thead{Time Till  \\ Infection} & \thead{Time  \\ Healthy} & \thead{Early \\ Detection} & \thead{Unsure \\ Day} \\ \hline
10 & 0 & 10 & NA & 0 & 0 \\ \hline
11 & 0 & 9 & NA & 0 & 0 \\ \hline
12 & 0 & 8 & NA & 0 & 0 \\ \hline
13 & 0 & 7 & NA & 1 & 0 \\ \hline
14 & 0 & 6 & NA & 1 & 0 \\ \hline
15 & 0 & 5 & NA & 1 & 0 \\ \hline
16 & 0 & 4 & NA & 1 & 0 \\ \hline
17 & 0 & 3 & NA & 1 & 0 \\ \hline
18 & 0 & 2 & NA & 1 & 0 \\ \hline
19 & 0 & 1 & NA & 1 & 0 \\ \hline
20 & 1 & 0 & 0 & 0 & 1 \\ \hline
21 & 0 & NA & 1 & 0 & 1 \\ \hline
22 & 0 & NA & 2 & 0 & 1 \\ \hline
23 & 0 & NA & 3 & 0 & 1 \\ \hline
24 & 0 & NA & 4 & 0 & 1 \\ \hline
25 & 0 & NA & 5 & 0 & 1 \\ \hline
26 & 0 & NA & 6 & 0 & 1 \\ \hline
27 & 0 & NA & 7 & 0 & 1 \\ \hline
\end{tabular}
\caption{AMA Structure}
\label{tab: AMA structure}
\end{center}

\end{table}

\subsection{Gradient Boosting Algorithm}
The algorithm used for this study was a Gradient Boosting Algorithm. This was chosen firstly due to it being in the family of Decision Tree algorithms which have already shown to work within a wide range of areas in the literature such as \cite{8610248}, \cite{CHANG2018914}. We then further decided to use a Gradient Boosting Algorithm due to the fact that it can discover many different types of patterns and results while still being less complex than other modelling types such as Multi-Layered Perceptron which allows in some way the end user to see which features were the most important in classifying the problem at hand.
\subsection{Class imbalance}
When there is a large class imbalance in a binary classification problem the algorithm being used may create results that are biased towards the class that is in the majority. There are many ways to overcome this with the two main contributors being \linebreak{} Resampling and also Threshold Moving.
\subsubsection{Resampling}
Resampling works on the idea of creating a more balanced data set of the different classes that are in your original data set. It can be split between Undersampling and Oversampling, within both areas there a wide range of different algorithims that have been constructed to reduce the effect of having unequal class sizes \cite{chawla_bowyer_hall_kegelmeyer_2002}, \cite{saez}, \cite{barua_islam_yao_murase_2014}, \cite{he_bai_garcia_li_2008} are just some examples. For this study we chose to use an Oversampling approach, this is due to the widely considered point that Undersampling has the capability to lose a large quantity of information from the majority class while also reducing the training set size to a much smaller size and thus over fitting to the minority class can occur. \par
\begin{figure}[ht]
\centering
\includegraphics[scale = 0.5]{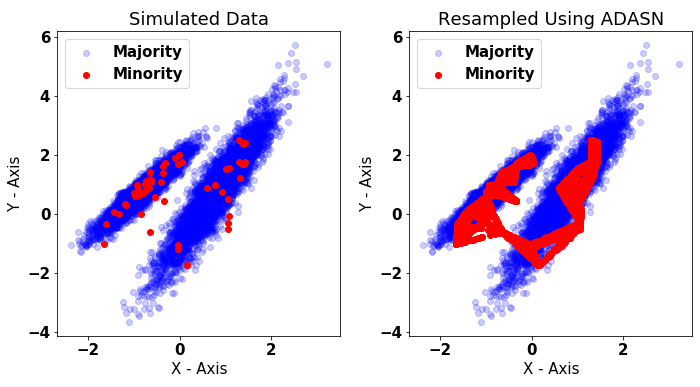}
\caption{Example of OverSampling}
\label{fig:Original}
\end{figure}

 Within the area of Oversampling there is a large amount of different algorithms that have been created to alter the class distribution. The main algorithim that is used in the literature is SMOTE \cite{chawla_bowyer_hall_kegelmeyer_2002} which creates new observations between two observations that are seen to be similar to each other. Its been shown that this algorithim has many drawbacks and as a result many new algorithims have come out that try to handle these problems, one such algorithim is ADASYN \cite{he_bai_garcia_li_2008} which tries to focus on creating new observations that are both similar to its own smaller class but also similar to the larger class. The general idea of oversampling can be seen in Figure \ref{fig:Original} which shows a simulated data set before and after ADASYN has been conducted.

\subsubsection{Threshold Moving}
Another way that the issue of Class Imbalance can be resolved is through the use of Threshold Moving. This method involves the use of moving the default 0.5 probability threshold to a more valid value. There are two main methods in which this can be accomplished which include the use of F1-score and Youden's j-Statistic \cite{youden_1950}. Moving the threshold by using a F1-score relies on calculating the final results for Recall/Sensitivity and Precision through the following equation:
\begin{equation}
    2 * \frac{Precision * Recall}{Precision + Recall}
\end{equation}
Moving the threshold by using Youden's j-Statistic is computed by the following \linebreak{} equation:
\begin{equation}
    sensitivity + specificity - 1
\end{equation}
The `best' threshold is seen as the probability threshold that results in the highest F1-score or Youden's j-Statistic for that respective feature.
For this study we chose to use Youden's j-Statistic to alter the threshold
\par Due to the fact that we are altering the probability threshold for this study we wanted to make sure that we weren't altering our results through the use of over fitting. To account for this we split our training data set into two separate distinct subsets. These were split to keep 80$\%$ of the data within the training set while\linebreak{} creating a smaller subset that could be used to calculate the Youden's j-Statistic and its threshold accompanying it. This threshold would then be used on the test set to make sure no over fitting was achieved.
\section{Results}
In this section we will look at how well both models do at their respective goal and how we measured the success of each model as this is different for both methods. \par As the data we are working with in this study was seen as imbalanced when we are reporting result metrics from our models we decided against the use of the more general metrics such as Accuracy, in favour of other metrics such as Specificity, Sensitivity, Balanced Accuracy and also Geometric Mean which are all described below. \begin{equation}
    Accuracy = \frac{TP + TN}{TP + TN + FP + FN}
\end{equation}
\begin{equation}
    Sensitivity = \frac{TP}{TP + FN}, \; Specificity = \frac{TN}{TN + FP}
\end{equation}
\begin{equation}
    Balanced \; Accuracy = \frac{Sensitivity + Specificity}{2}
\end{equation}
\begin{equation}
    Geometric \; Mean = \sqrt{Sensitivity * Specificity}
\end{equation}The reason for this is due to the fact that in the creation of Accuracy the overall imbalance of a data set is not taken into account thus with a dataset that has 90\% of the observations belonging to the larger class we could achieve an overall accuracy value of 90\% by simply predicting all observations as the larger class, this while giving us a high value for accuracy is pointless in the modeling of the smaller class as that is normally the class of interest.
\subsection{SMA}
The output from Model 1 is the easiest to interpret as we can simply look at the confusion matrix which is seen in Table~\ref{tab:Model 1}. The 4 different outputs of Table \ref{tab:Model 1} can be seen as the class label that will be given the next time the cow gets tested for SCC. We can see the final results in Tables \ref{tab:Model 1 SCM},\ref{tab:Model 1 CM} and then also the Statistical Measures in Table~\ref{tab:Model 1 Results}. We can see from these three tables that this model structure is able to classify well for both cows that are going to be healthy or going to have a high SCC the next time it gets recorded.

\begin{table}[ht]
\begin{center}
\begin{tabular}{c|c|c|c|}
\cline{3-4}
\multicolumn{2}{c|}{\multirow{2}{*}{}}                            & \multicolumn{2}{c|}{\textbf{Predicted}} \\ \cline{3-4} 
\multicolumn{2}{c|}{}                                             & Negative           & Positive          \\ \hline
\multicolumn{1}{|c|}{\multirow{2}{*}{\textbf{Actual}}} & Negative  & Healthy Milking Found              & Healthy Milking not Found               \\ \cline{2-4} 
\multicolumn{1}{|c|}{}                                 & Positive & Infected Milking not Found                & Infected Milking Found               \\ \hline
\end{tabular}
\caption{\label{tab:Model 1}SMA Classification Matrix}
\end{center}
\end{table}

\begin{table}[ht]
\begin{center}
    
\begin{tabular}{c|c|c|c|}
\cline{3-4}
\multicolumn{2}{c|}{\multirow{2}{*}{}}                            & \multicolumn{2}{c|}{\textbf{Predicted}} \\ \cline{3-4} 
\multicolumn{2}{c|}{}                                             & Healthy            & Infected           \\ \hline
\multicolumn{1}{|c|}{\multirow{2}{*}{\textbf{Actual}}} & Healthy  & 20462              & 2608               \\ \cline{2-4} 
\multicolumn{1}{|c|}{}                                 & Infected & 671                & 2092               \\ \hline
\end{tabular}
\caption{\label{tab:Model 1 SCM}Model 1 Classification Matrix using SCM records}
\end{center}
\end{table}

\begin{table}[ht]
\begin{center}
    
\begin{tabular}{c|c|c|c|}
\cline{3-4}
\multicolumn{2}{c|}{\multirow{2}{*}{}}                            & \multicolumn{2}{c|}{\textbf{Predicted}} \\ \cline{3-4} 
\multicolumn{2}{c|}{}                                             & Healthy            & Infected           \\ \hline
\multicolumn{1}{|c|}{\multirow{2}{*}{\textbf{Actual}}} & Healthy  & 20031              & 3039               \\ \cline{2-4} 
\multicolumn{1}{|c|}{}                                 & Infected & 602                & 2161        \\ \hline
\end{tabular}
\caption{\label{tab:Model 1 CM}Model 1 Classification Matrix using CM records}
\end{center}
\end{table}

\begin{table}[ht]
\begin{center}
\begin{tabular}{c|cc|}
\cline{2-3}
                                        & \multicolumn{2}{c|}{\textbf{Results}} \\ \cline{2-3} 
                                        & \multicolumn{1}{c|}{SCM records}   & CM records   \\ \hline
\multicolumn{1}{|c|}{Balanced Accuracy} & \multicolumn{1}{c|}{82.21}         & 82.52        \\ \hline
\multicolumn{1}{|c|}{Geometric Mean}    & \multicolumn{1}{c|}{81.95}       & 82.41         \\ \hline
\multicolumn{1}{|c|}{Specificity}       & \multicolumn{1}{c|}{88.7}         & 86.83        \\ \hline
\multicolumn{1}{|c|}{Sensitivity}       & \multicolumn{1}{c|}{75.71}         & 78.21        \\ \hline
\end{tabular}
\caption{\label{tab:Model 1 Results}Model 1 Accuracy Metrics}
\end{center}
\end{table}

\subsection{AMA}
It is possible to look at the results for this approach in the same way that is given for the first modelling approach which is given in Tables \ref{tab:SCM using CM},\ref{tab:CM using CM},\ref{tab:SCM using SCM} and \ref{tab:CM using SCM}. Yet when we look at these tables more closely we can see that the results given by these aren't exactly what we want to showcase, to accomplish this we must alter the final results in a small way. \par To interpret the results outputted from AMA requires more preparation as each milking that is predicted doesn't necessarily need to be used within the final output of results. The idea behind this is due to the fact that we are creating alerts that would inform the farmer whether or not the particular cow is at risk of a mastitis infection within the next 7 days. Thus if an alert was given when the cow is 20 Days in Milk that said the cow is at risk and another alert was given at 21 Days in Milk the farmer wouldn't need the results from the later milking due to the farmer already being able to take the neccessary precautions. Thus we altered the idea of \emph{Specificity} to take this into account. This was done by first calculating the total amount of infections that were included in the final test dataset and then after the individual alerts were created we illustrated a infection being found if it had atleast 1 alert created within the 7 days before hand. This lead us to further be able to calculate the earliest time at which each infection was found and as a result use that as a way to track how the progress of infections were found as we got closer to the days of the infections. \par Further improvements were also made to accomodate the alerts created for \linebreak{} healthy milkings (or those milkings that were predicted not to have any infections within the next 7 days) throughout the year instead of just one calculated result. This allowed us to be able to see how our model predicted healthy outcomes as the lactation period continued. \par As with this model we were able to predict for both SCM and CM infections we were left with two seperate results for both the modified \emph{Sensitivity} and also \emph{Specificity}. 

\begin{table}[ht]
\begin{center}
    
\begin{tabular}{c|c|c|c|}
\cline{3-4}
\multicolumn{2}{c|}{\multirow{2}{*}{}}                            & \multicolumn{2}{c|}{\textbf{Predicted}} \\ \cline{3-4} 
\multicolumn{2}{c|}{}                                             & Healthy            & Infected           \\ \hline
\multicolumn{1}{|c|}{\multirow{2}{*}{\textbf{Actual}}} & Healthy  & 156602              & 30048               \\ \cline{2-4} 
\multicolumn{1}{|c|}{}                                 & Infected & 2965                & 6194        \\ \hline
\end{tabular}
\caption{\label{tab:SCM using CM}Model 2 Classification Matrix for SCM model using CM records}
\end{center}
\end{table}

\begin{table}[ht]
\begin{center}
    
\begin{tabular}{c|c|c|c|}
\cline{3-4}
\multicolumn{2}{c|}{\multirow{2}{*}{}}                            & \multicolumn{2}{c|}{\textbf{Predicted}} \\ \cline{3-4} 
\multicolumn{2}{c|}{}                                             & Healthy            & Infected           \\ \hline
\multicolumn{1}{|c|}{\multirow{2}{*}{\textbf{Actual}}} & Healthy  & 193074              & 19575               \\ \cline{2-4} 
\multicolumn{1}{|c|}{}                                 & Infected & 1030                & 906        \\ \hline
\end{tabular}
\caption{\label{tab:CM using CM}Model 2 Classification Matrix for CM model using CM records}
\end{center}
\end{table}

\begin{table}[ht]
\begin{center}
    
\begin{tabular}{c|c|c|c|}
\cline{3-4}
\multicolumn{2}{c|}{\multirow{2}{*}{}}                            & \multicolumn{2}{c|}{\textbf{Predicted}} \\ \cline{3-4} 
\multicolumn{2}{c|}{}                                             & Healthy            & Infected           \\ \hline
\multicolumn{1}{|c|}{\multirow{2}{*}{\textbf{Actual}}} & Healthy  & 151097              & 35553               \\ \cline{2-4} 
\multicolumn{1}{|c|}{}                                 & Infected & 2598                & 6561        \\ \hline
\end{tabular}
\caption{\label{tab:SCM using SCM}Model 2 Classification Matrix for SCM model using SCM records}
\end{center}
\end{table}

\begin{table}[ht]
\begin{center}
    
\begin{tabular}{c|c|c|c|}
\cline{3-4}
\multicolumn{2}{c|}{\multirow{2}{*}{}}                            & \multicolumn{2}{c|}{\textbf{Predicted}} \\ \cline{3-4} 
\multicolumn{2}{c|}{}                                             & Healthy            & Infected           \\ \hline
\multicolumn{1}{|c|}{\multirow{2}{*}{\textbf{Actual}}} & Healthy  & 189893              & 22756               \\ \cline{2-4} 
\multicolumn{1}{|c|}{}                                 & Infected & 1023                & 913        \\ \hline
\end{tabular}
\caption{\label{tab:CM using SCM}Model 2 Classification Matrix for CM model using SCM records}
\end{center}
\end{table}

\subsubsection{Sub Clinical Infections}
\begin{figure}[ht]
\centering
\includegraphics[scale = 0.5]{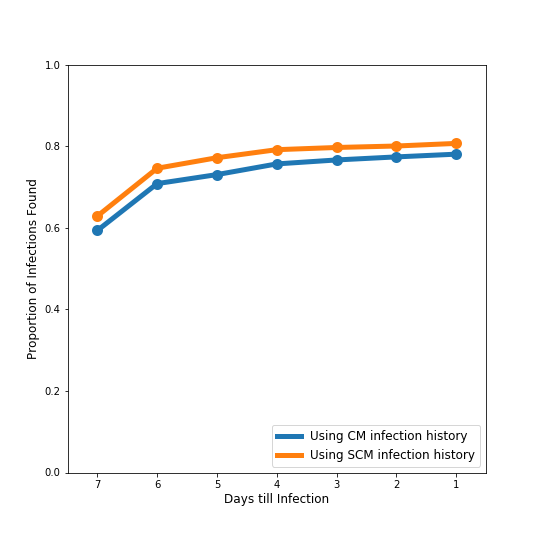}
\caption{Timeline of SCM infections}
\label{fig:Timeline SCM}
\end{figure}

The final results for predicting SCM infections is shown in Table \ref{tab:sub_clinical method 2}, \ref{tab:Healthy SCM} and \linebreak{} also Figure \ref{fig:Timeline SCM} which illustrates the progress of picking up infections as the timeline progresses and then also the average proportion of healthy milkings correctly \linebreak{} predicted each day of the year using both SCM and CM infection history separately.

\begin{table}[ht]
\begin{center}
\begin{tabular}{@{}llllllll@{}}
\toprule
Days till Infection        & 7     & 6    & 5     & 4     & 3     & 2     & 1     \\ \midrule
Using Sub Clinical Records & 62.90 & 74.63 & 77.21 & 79.18 & 79.72 & 80.08 & 80.72 \\
Using Clinical Records     & 59.37 & 70.86 & 73.07 & 75.69 & 76.66 & 77.41 & 78.08 \\ \bottomrule
\end{tabular}
\caption{Sub-Clinical Infections}
\label{tab:sub_clinical method 2}
\end{center}
\end{table}

\begin{table}[ht]
\begin{center}
\begin{tabular}{@{}lll@{}}
\toprule
                                           & SCM History & CM History \\ \midrule
Proportion of Healthy Correctly Classified & 81.19       & 84.51      \\ \bottomrule
\end{tabular}
\caption{Healthy Milkings for SCM model}
\label{tab:Healthy SCM}
\end{center}
\end{table}

\subsubsection{Clinical Infection} 
One addition was included within the models when trying to predict CM, this is to do with the idea of how many days for each milking since SCC was last recorded. This was not included within the SCM predicting models due to the idea that SCM can only be found when SCC is measured and due to it being measured on average 7 days apart we thought it might lead to overfitting which is not present when measuring for CM due to them being found whenever the farmer notices them instead. \par
\begin{figure}[!t]
\centering
\includegraphics[scale = 0.5]{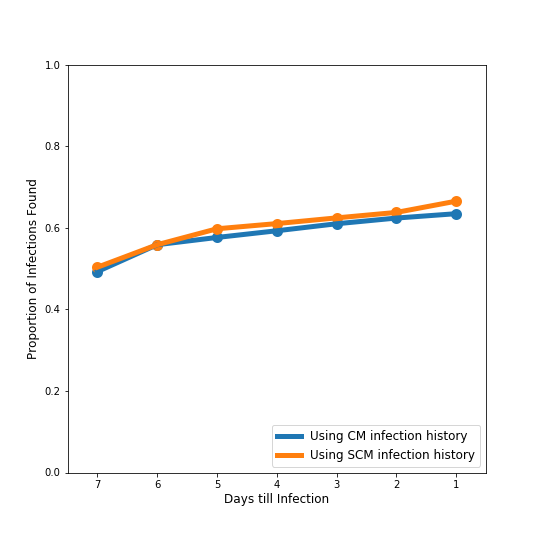}
\label{fig:Timeline CM}
\caption{Timeline of CM infections}
\end{figure}

 The final results for predicting CM infections is shown in Table \ref{tab:clinical method 2}, \ref{tab:Healthy CM} and also Figure \ref{fig:Timeline CM} which illustrates the progress of picking up infections as the timeline progresses and then also the average proportion of healthy milkings correctly \linebreak{}predicted each day of the year using both SCM and CM infection history separately. 

\begin{table}[ht]
\begin{center}
    
\begin{tabular}{@{}lll@{}}
\toprule
                                           & SCM History & CM History \\ \midrule
Proportion of Healthy Correctly Classified & 88.02       & 89.98      \\ \bottomrule
\end{tabular}
\caption{Healthy Milkings for CM model}
\label{tab:Healthy CM}
\end{center}
\end{table}

\begin{table}[ht]
\begin{center}
\begin{tabular}{@{}llllllll@{}}
\toprule
Days till Infection        & 7     & 6     & 5     & 4    & 3     & 2     & 1     \\ \midrule
Using Sub Clinical Records & 50.35 & 55.87 & 59.78 & 61.05 & 62.45 & 63.79 & 66.55 \\
Using Clinical Records     & 49.28 & 55.87 & 57.65 & 59.29 & 61.05 & 62.41 & 63.48 \\ \bottomrule
\end{tabular}
\caption{Clinical Infections}
\label{tab:clinical method 2}
\end{center}
\end{table}

\section{Discussion}
When we look at the results we can clearly see the main difference between SMA and AMA with regards to its capability of determining SCM infections is the fact that SMA achieves a much lower rate of alerts that are said to be Positive when in fact the cow would be healthy which is seen as the total amount of FN. Yet we can also see that the Sensitivity value for SMA is higher than that for AMA in the majority of cases. This leads us to the opinion that the inclusion of many other days reduces the effect of classifying an infection further out from its onset.  This might be as a result of SCC only being recorded every 7 days alongside the majority of other features and thus with AMA the inclusion of the Time Series adapted feature set leads to the introduction of many rows that apart from the yield and maximum flow rate values are exactly the same for 3 days in a row. This could potentially be a reason for why there is not an increase when there is more information available for our model to work on. Thus it might be appropriate to think that our model could work better if each milking feature was recorded more frequently than the average of 7 days in this study. This would also allow us to use a smaller period to define the Time Series features on as the values of 15 and 30 days were chosen to make sure that these features were being constructed on more than just one unique value.
\par We can also see that the act of correctly finding CM is much harder than that of finding SCM, in part this could come down to the fact that there was a much larger proportion of SCM infections within our data set in the first place but also CM infections can come in many different forms mainly being Contagious or  \linebreak{}Environmental infections \cite{hyde_down}. It has been show that these two subsets of mastitis infections while still being mastitis occur in different ways but also are illustrated by different results which could have resulted in our CM infections grouping themselves into so called 'small disjoints' \cite{jo_japkowicz_2004} which is a major issue within imbalanced data. This occurs when the minority class is actually made up of a multitude of smaller classes which might belong in some way to the same class but will still have large differences to other observations in that class. This problem seems to only come about when you deal with a small sample size and in most instances will disappear as the positive samples get bigger.
\section{Conclusion}
Within this study we tried to answer two separate questions, the first being if we could predict SCM infections one SCC recording in advance and the second was to see if we could predict SCM and CM infections at least 7 days before they would be detectable. From our results we can say that this has been accomplished with a relatively high accuracy for both questions. Yet even though the results were relatively high there still lies areas that that the results could be improved upon. A main aspect of this is with regards to the large amounts of FN created. This was reduced when we go to the SMA model which only looked at days where SCC was recorded. Yet the advantages of AMA which include the ability to classify CM infections might outweigh the disadvantage of having high false alerts created.

\section*{Acknowledgements}
This publication has emanated from research conducted with the financial support of Science Foundation Ireland (SFI) and the Department of Agriculture, Food and Marine on behalf of the Government of Ireland under Grant Number [16/RC/3835].
\bibliographystyle{plain}
\bibliography{references}
\end{document}